%% file: main.tex
\documentclass[sigconf,9pt]{acmart}
\acmConference[DAC '26]{Design Automation Conference}{June 23--27, 2024}{San Francisco, CA}

\usepackage{multirow}
\usepackage{subfigure}
\usepackage{bbding}
\usepackage{mathtools}

\usepackage{algorithm}
\usepackage{algpseudocode}
\algtext*{EndFor}
\algtext*{EndFunction}
\algtext*{Until}
\algtext*{EndIf}
\algtext*{EndWhile}
\usepackage{amsmath}
\usepackage{array}
\usepackage[referable]{threeparttablex}
\usepackage{tablefootnote}

\usepackage{tikz}
\usepackage{xcolor}
\usepackage{float}
\usepackage{graphicx} 
\usepackage{caption} 
\usepackage{subcaption}
\usepackage{colortbl}
\usepackage{enumitem}

\AtBeginDocument{%
  \providecommand\BibTeX{{%
    \normalfont B\kern-0.5em{\scshape i\kern-0.25em b}\kern-0.8em\TeX}}}

\copyrightyear{2026}
\acmYear{2026}
\setcopyright{cc}
\setcctype{by}
\acmConference[DAC '26]{63rd ACM/IEEE Design Automation Conference}{July 26--29, 2026}{Long Beach, CA, USA}
\acmBooktitle{63rd ACM/IEEE Design Automation Conference (DAC '26), July 26--29, 2026, Long Beach, CA, USA}
\acmDOI{10.1145/3770743.3804165}
\acmISBN{979-8-4007-2254-7/2026/07}

\definecolor{lightgreen}{RGB}{198, 224, 183}

\begin{document}

\input{_txt/abstract}


\title{G-Power: Architecture-level GPU Power Modeling with Aggregated Knowledge Foundations from Known GPUs}



\author{Qijun Zhang$^1$, Yao Lu$^1$, Shang Liu$^1$, Mengming Li$^1$, Chen Zhang$^2$, Dongbo Wang$^3$, Zhiyao Xie$^1$}
\authornote{Corresponding Author}
\affiliation{%
  \institution{$^1$Hong Kong University of Science and Technology, $^2$Shanghai Jiao Tong University, $^3$UniVista}
  \country{\{qzhangcs, yludf, sliudx, mengming.li\}@connect.ust.hk, chenzhang.sjtu@sjtu.edu.cn, wdb@univista-isg.com}
  \country{eezhiyao@ust.hk}
}

\maketitle

\input{_txt/1_introduction}
\input{_txt/3_method}

\input{_txt/4_result}
\input{_txt/7_conclusion}

\bibliographystyle{ACM-Reference-Format}
\bibliography{refs}

\end{document}

%% file: _txt/abstract.tex
\begin{abstract}
Graphics Processing Units (GPUs) have been serving as critical computation resources for large-scale parallel computations. With increasing chip complexity, power efficiency has become an important design objective for modern GPUs. GPU power optimization relies on fast power evaluation, requiring architecture-level GPU power model. However, because of the time-consuming power label collection, only simple microbenchmarks are adopted for training. The limitation of microbenchmarks as training data incurs low accuracy for existing architecture-level GPU power models like AccelWattch.\looseness=-1

To address the limitation of microbenchmarks as training data, we propose G-Power, an architecture-level GPU power modeling framework that utilizes additional known GPU chips to provide additional knowledge. G-Power utilizes the aggregated knowledge foundation from additional known GPU chips and then performs fine-tuning on our target GPU. To provide foundations with additional known GPU chips and capture the similarity to utilize these foundations for fine-tuning, G-Power adopts a three-phase algorithm consisting of 1) pre-training with additional known chips, 2) attention-inspired aggregation, and 3) fine-tuning on our target GPU. We evaluate G-Power on four modern NVIDIA GPUs, demonstrating high accuracy. G-Power can achieve a low MAPE of 14\% and a high correlation coefficient $R$ of 0.88 on average, which are 22\% lower MAPE and 0.36 higher $R$ than AccelWattch.

\end{abstract}

%% file: _txt/1_introduction.tex
\section{Introduction}
\label{sec:intro}

Graphics Processing Units (GPUs) have become critical computation resources for accelerating large-scale parallel computations, such as machine learning, big data, and scientific computing~\cite{krizhevsky2012imagenet, top500, cloudai, hpc}. With the increasing complexity of GPU chips, power efficiency is becoming an increasingly important design objective for modern GPU designs, where even consumer-grade GPUs like RTX3090 can reach 350W. 
Power optimization of GPUs relies on power evaluation. Standard power simulation flow goes through RTL simulation, logic synthesis, physical design, and power simulation~\cite{design-compilier,vcs,ptpx}. 
Such a power simulation flow is time-consuming, motivating architecture-level power models. 

\begin{figure}[!t]
\centering
\vspace{-.1in}
\includegraphics[width=0.47\textwidth]{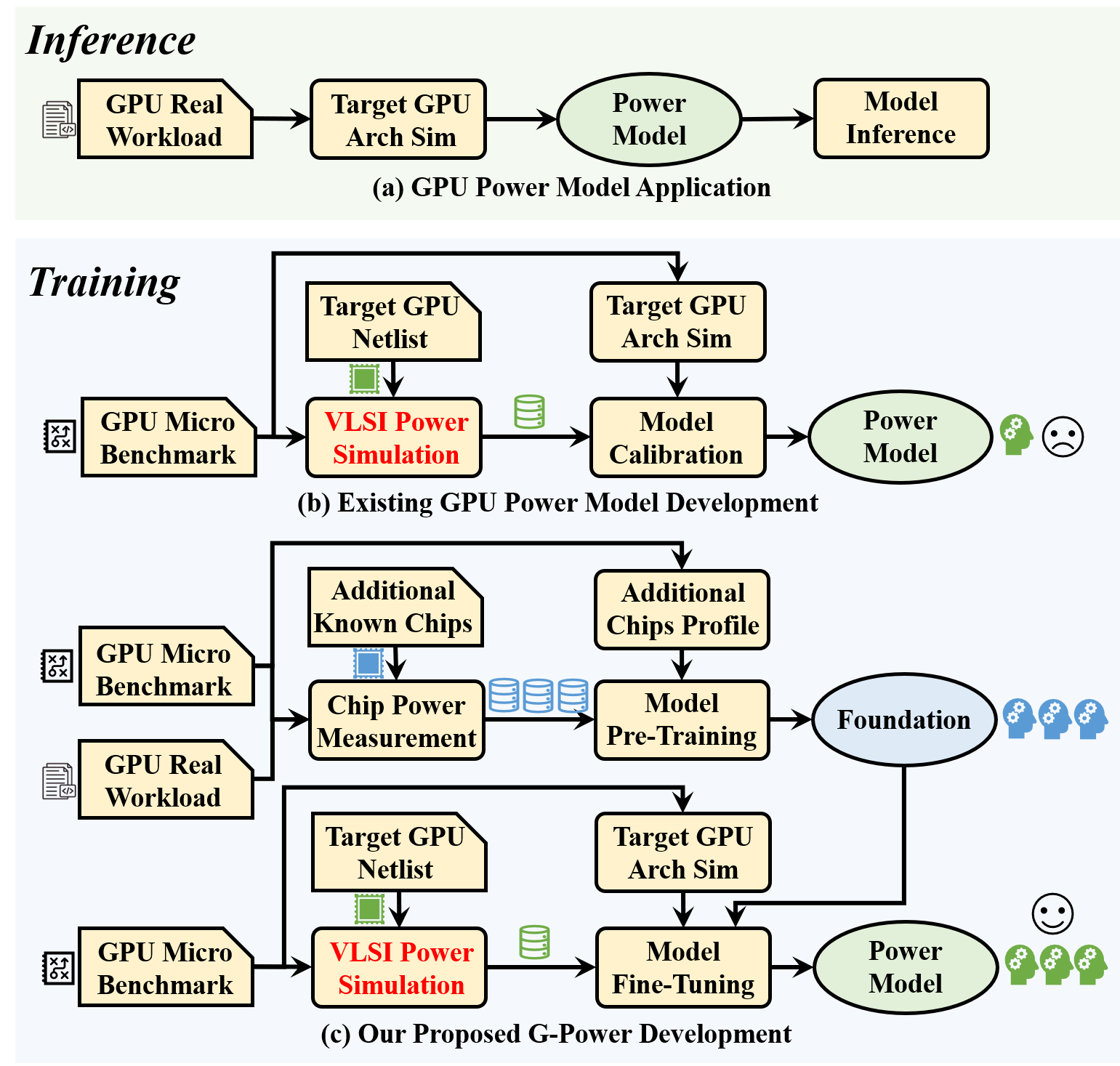}
\vspace{-.05in}
\caption{Workflow comparison between existing GPU power modeling~\cite{kandiah2021accelwattch} and our proposed G-Power. G-Power learns from additional known chips through the pre-training foundation to enable accurate modeling for our target GPU.}
\vspace{-.15in}
\label{overview}
\end{figure}

\textbf{Power Model:} 
Recent general RTL-stage power models~\cite{xie2021apollo,xie2022deep,zhou2019primal,kim2019simmani} eliminate the requirement of the VLSI flow, but RTL simulation is also prohibitively time-consuming. 
There have also been many architecture-level power models for CPU~\cite{li2009mcpat,brooks2000wattch,zhang2023panda,zhang2025firepower,zhang2025autopower,zhangarchpower,zhang2026readypower,zhang2024architecture}, while GPU models~\cite{kandiah2021accelwattch,leng2013gpuwattch} are few. GPU~\cite{kandiah2021accelwattch,leng2013gpuwattch}. GPU power modeling is more challenging than CPUs because the lack of high-quality open-source GPU designs forces researchers to use real GPU chips for power label collections, where no intermediate data is available.

\textbf{Architecture-level GPU Power Models:} AccelWattch~\cite{kandiah2021accelwattch}, the updated version of GPUWattch~\cite{leng2013gpuwattch}, has been widely adopted for GPU power modeling at the architecture level. 
As shown in Figure~\ref{overview}(a), for our target GPU design, AccelWattch performs architect-ure-level performance simulation for workloads to generate architectural events, and then takes these events to estimate GPU power. 
The development flow of AccelWattch is illustrated in Figure~\ref{overview}(b). For each GPU design, the AccelWattch is calibrated with linear regression. 
Because of the time-consuming power label collection, only short microbenchmarks with hundreds of instructions are adopted for calibration. After calibration, the model can predict the power consumption of long real workloads that have millions of instructions with fast architectural simulation. 
However, our measurement shows that when applying AccelWattch to Ampere and Ada GPUs, the mean absolute percentage error (MAPE) can be over 30\%, and the correlation $R$ can be under 0.6. Such a low accuracy is unacceptable to guide GPU power optimization.

\textbf{Limitation of Microbenchmarks as Training Data:} 
The low accuracy of the existing GPU power model is due to limitations of microbenchmarks as training data: the microbenchmarks' data distribution of the feature and label is different from real workloads. These microbenchmarks are short and simple, demonstrating a different data distribution from the long and complex real workloads that can not be adopted for training because of the time-consuming power simulation for our target GPU. This distribution difference results in low accuracy of the existing GPU power model workflow.

\textbf{Our Solution--Use Data of Additional Known Chips:} To address the limitation above, we propose G-Power, an architec-ture-level GPU power model that learns from additional known GPU chips with a pre-training foundation. 
Compared with the existing GPU power model that is trained with only our target GPU, G-Power utilizes additional known GPU chips to provide additional knowledge. 
Figure~\ref{overview}(c) illustrates the development flow of our proposed G-Power framework. 
G-Power learns from multiple additional known GPU chips, where data from both microbenchmarks and real workloads are available because of fast real chip power measurement. 

\textbf{How to Use Additional Known Chips:} We point out that there are both similarities and differences among different GPUs. 
The similarity exists in both overall architecture and pipeline microarchitecture: 1) GPUs adopt a similar architectural design paradigm with SM, NoC, and memory partition. 2) GPUs have a similar pipeline design following the same hyper-threading microarchitecture that hides memory access latency. 
However, GPUs also have differences: 1) Different GPUs have different hardware configurations that are unavailable, such as the width of each pipeline stage, the number of processing units, and the cache configurations. 2) Different GPUs have different microarchitecture designs. For example, the warp scheduler, tensor core design, and memory access unit may differ across GPUs. 
Therefore, the key to G-Power is to \emph{provide foundations with additional known GPU chips and capture the similarity to utilize these foundations for fine-tuning}.

G-Power adopts a three-phase algorithm: 
1) \emph{Pre-training with additional known chips}. It builds one foundation model for each additional known GPU chip. The foundation model is trained with both the microbenchmarks and the real workloads, incorporating comprehensive knowledge across programs. 2) \emph{Attention-inspired aggregation}. This step captures the similarity. It estimates the similarity and then aggregates these foundation models based on this similarity to generate an aggregated foundation. 3) \emph{Fine-tuning on our target GPU}. It fine-tunes the aggregated foundation for our target GPU with microbenchmark data from our target GPU. 

\textbf{Contributions:} Our contributions are summarized below.
\begin{itemize}[leftmargin=0.4cm]
    \item We propose G-Power, an architecture-level GPU power modeling framework that utilizes additional known chips to provide additional knowledge.
    \item G-Power introduces a three-phase algorithm to provide foundations with additional known GPU chips and capture the similarity to utilize these foundations for fine-tuning. It includes pre-training with additional known chips, attention-inspired aggregation, and fine-tuning on our target GPU.  
    \item Evaluation of G-Power on four modern NVIDIA GPUs shows that G-Power can achieve a low MAPE of 14\% and a high correlation $R$ of 0.88 on average, which are 22\% lower MAPE and 0.36 higher $R$ than AccelWattch. 
\end{itemize}

%% file: _txt/3_method.tex
\section{G-Power Overview}
\label{sec:overview}

\begin{table}[!t]
      \centering
      \renewcommand{\arraystretch}{1.1}
      \resizebox{0.44\textwidth}{!}{
        \begin{tabular}{ |c|c|c| } 
        \hline
        Domain & \multicolumn{2}{c|}{Architectural Event} \\
        \hline
        \hline
         \multirow{10}{*}{SM Array} & Total PTX Instruction & Total SASS Instruction \\
         \cline{2-3}
         & Float and Integer Instruction & Fp64 Instruction \\
         \cline{2-3}
         & DCache Read Hit & DCache Read Miss \\
         \cline{2-3}
         & DCache Write Hit & DCache Write Miss \\
         \cline{2-3}
         & Constant Cache Hit & Constant Cache Miss \\
         \cline{2-3}
         & Shared Memory Access & Textural Cache Access \\
         \cline{2-3}
         & ALU Access & Tensor Core Access \\
         \cline{2-3}
         & FMA Access & DFMA Access \\
         \cline{2-3}
         & DPU Access & EXU Access \\
         \cline{2-3}
         & SM Activity & Warp Activity \\
        \hline
        \multirow{3}{*}{NoC \& L2Cache} & L2Cache Read Hit & L2Cache Read Miss \\
         \cline{2-3}
         & L2Cache Write Hit & L2Cache Write Miss \\
         \cline{2-3}
         & NoC Request & NoC Response \\
         \hline 
         Global Memory & Memory Read & Memory Write \\
         \hline 
        
        \end{tabular}
        }
        \caption{Our adopted architectural events in G-Power. It includes 28 events from the three GPU domains.}
        \vspace{-.35in}
        \label{tbl:event}
\end{table}

In the GPU, each component has its architectural events that trigger the activity of its internal circuit. Because the same event always triggers the same circuit part, we can formulate the GPU power consumption as a linear function of architectural events. Denoting the number of events as $N$, the maximum power of component $i$ as $p_{i}$, the idle power as $p_{idle}$, the $i$-th architectural event normalized by execution time as $e_{i}$, GPU power is formulated in Equation~\ref{eq:power1}.
\begin{align}
P(\{e_i\}) & = \sum_{i=1}^{N} e_{i}*p_{i} + p_{\text{idle}} \label{eq:power1}
\end{align}
where architectural events $e_i$ are model inputs and the $p_{i}$ and $p_{idle}$ are model parameters. 
In G-Power, we adopt 28 architectural events $e_i$ ($N=28$), covering important components of the GPU. These architectural events are listed in Table~\ref{tbl:event}. These events are classified by GPU domains, including Streaming Multiprocessors (SMs), on-chip network (NoC) and L2 Cache, and global memory, where components within a domain are tightly-coupled. 

\begin{figure*}[!t]
\centering
\includegraphics[width=0.97\textwidth]{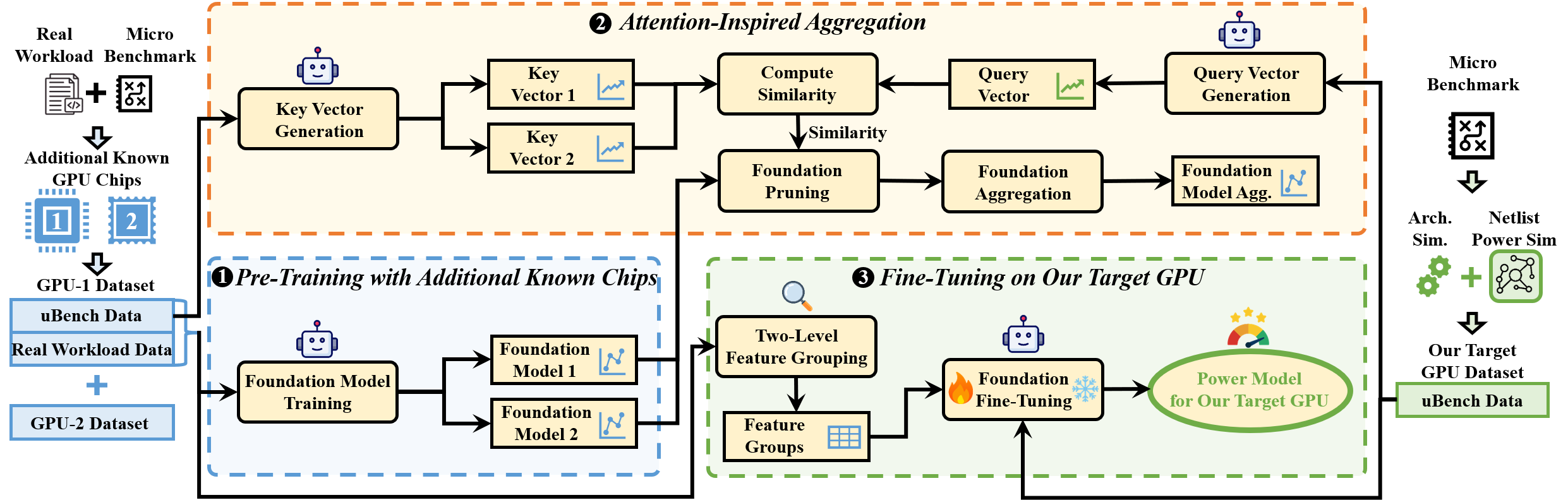}
\vspace{-.12in}
\caption{The three-phase framework of G-Power. 1) Pre-training with additional known chips. Multiple foundation models are trained based on datasets collected from multiple additional known GPU chips, where the dataset includes both the microbenchmark and real workload data. 2) Attention-inspired aggregation. Key and query vectors are generated based on microbenchmark data of additional known GPU chips and our target GPU, respectively. Foundations are aggregated based on similarities between the query vector and each key vector. 3) Fine-tuning on our target GPU. Features are grouped with two-level feature grouping, and the aggregated foundation is fine-tuned, where parameter ratios within each group are frozen.}
\vspace{-.15in}
\label{framework}
\end{figure*}

Figure~\ref{framework} illustrates the overview of our proposed G-Power, a methodology designed to enhance power modeling for our target GPU by leveraging additional known chip data. G-Power consists of three phases: pre-training with additional known chips, attention-inspired aggregation, and fine-tuning on our target GPU.

\textbf{1. Pre-training with additional known chips.} Multiple foundation models are trained using datasets collected from a variety of additional known GPU chips. These datasets include both carefully designed microbenchmarks and real workload traces. A key advantage of this step is its ability to incorporate rich and diverse power characteristics from known GPU designs, thereby establishing a foundation that encapsulates the power behavior of additional known GPU chips. This provides additional knowledge that would otherwise be unavailable when modeling our target GPU. \looseness=-1

\textbf{2. Attention-inspired aggregation.} Key and query vectors are generated from the microbenchmark data of additional known GPUs and our target GPU, respectively. The foundation models are then aggregated by evaluating the similarity between the target’s query vector and each additional key vector. The strength of this approach lies in its deliberate and aligned comparison mechanism: since both the additional and target GPUs can use the same microbenchmarks to generate data, the similarity is computed in a consistent, aligned feature space. This ensures that the aggregation meaningfully captures architectural similarities, effectively transferring the most relevant knowledge to our target GPU. 

\textbf{3. Fine-tuning on our target GPU.} In the final phase, we first group tightly coupled features via a two-level feature grouping strategy. The aggregated foundation model is then fine-tuned while keeping the parameter ratios within each feature group frozen. This design directly addresses the challenge of limited target-side data. By structurally constraining the tunable parameters, we effectively reduce the risk of overfitting, enabling stable adaptation to our target GPU. As a result, the knowledge encapsulated in pre-trained foundations is specialized to our target GPU in a robust way.

\section{Methodology}

We introduce our proposed three-phase algorithm in G-Power framework in this section. We describe pre-training with additional known chips in Section~\ref{sec:pt}, attention-inspired aggregation in Section~\ref{sec:agg}, and fine-tuning on our target GPU in Section~\ref{sec:ft}.

\subsection{Pre-Training with Additional Known Chips}
\label{sec:pt}

This phase trains the foundation model for each additional known GPU chip. As discussed in Section~\ref{sec:overview}, G-Power formulates the GPU power consumption as a linear function of architectural events. Therefore, we perform the linear regression for each chip dataset to build the foundation model for each additional known chip. 
Each chip dataset includes architectural events and power labels for both the microbenchmarks and real workloads, incorporating comprehensive knowledge about the GPU chip's power characteristics.

\subsection{Attention-Inspired Aggregation}
\label{sec:agg}

We propose attention-inspired aggregation to capture the similarity between our target GPU and each additional known GPU chip to utilize foundations wisely. Inspired by the attention~\cite{vaswani2017attention}, we represent each additional known chip as a key vector and our target GPU as the query vector, and then aggregate all foundation models based on similarity measured with query vector and key vector. \looseness=-1

\subsubsection{Key and Query Vector Generation}

A key vector is generated for each additional known GPU chip, which is paired with its foundation model. The query vector is generated for our target GPU. Our insight is that the coefficients of the linear model reflect the characteristics of the power model. Therefore, we perform the linear regression on each additional known chip and our target GPU, and collect coefficients as key and query vectors, where the coefficients are normalized to guarantee these vectors are unit vectors. 

Different from the foundation model that is trained on both the microbenchmark and real workload data, both the key and query vectors are trained on only the microbenchmark data. The reason is below: Only microbenchmark data is available to our target GPU, so the query vector can only be trained with microbenchmark data. For the key vector, it should be an equivalent counterpart to the query vector, so the key vector is also trained with only the microbenchmark data. Otherwise, the distribution difference between the microbenchmark and the real workload data will generate a biased vector that is not a counterpart to the query vector.

\subsubsection{Foundation Pruning and Aggregation}

\begin{figure}[!t]
\centering
\includegraphics[width=0.46\textwidth]{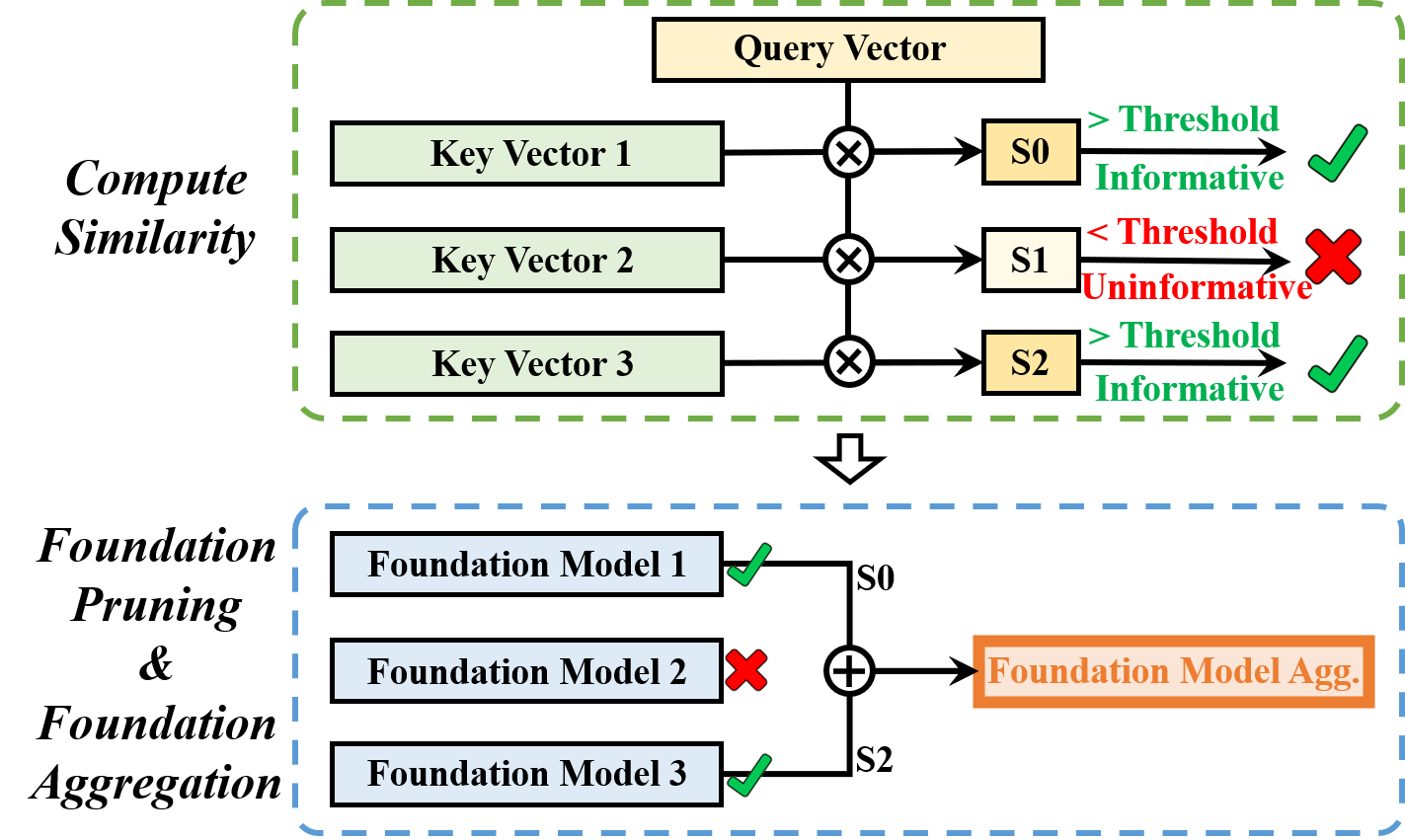}
\vspace{-.15in}
\caption{Illustration of attention-inspired aggregation. Each key vector is compared with the query vector to select informative foundations. All informative foundations are aggregated with similarities as weights.}
\vspace{-.2in}
\label{fondationsyn}
\end{figure}

With the key and query vectors generated, we aggregate all foundation models from additional known GPU chips based on similarities computed between the query vector and all key vectors. We first prune all foundation models with low similarities, and then aggregate the remaining foundation models by weighted-averaging them with similarities as weights. This enables G-Power to utilize the foundations wisely based on similarity.

The detail is illustrated in Figure~\ref{fondationsyn}. First, the query vector is compared with key vectors to compute similarities. The cosine similarities between the query vector of our target GPU and all key vectors of additional known GPU chips are calculated. 
With a pre-defined threshold, these foundations are classified as informative foundations~\cite{li2022transfer} and uninformative foundations, based on whether the similarity is larger than the threshold or not. The informative foundation is a foundation with high similarity, indicating that this additional known GPU is similar to our target GPU, so the knowledge of its foundation is likely to be useful. Conversely, the uninformative foundation is a foundation with low similarity, where the additional known GPU is different from our target GPU, and the knowledge may not be useful for our target GPU. 

Second, foundation pruning and foundation aggregation are performed with the identified informative foundations and calculated similarities to generate the final aggregated foundation model. The uninformative foundations are pruned because the knowledge of these may not be useful. Then, all informative foundations are weight-averaged to generate the final aggregated foundation model, where the weights are similarities normalized with softmax, similar to the operation in the attention mechanism.

\begin{figure}[!t]
\centering
\includegraphics[width=0.45\textwidth]{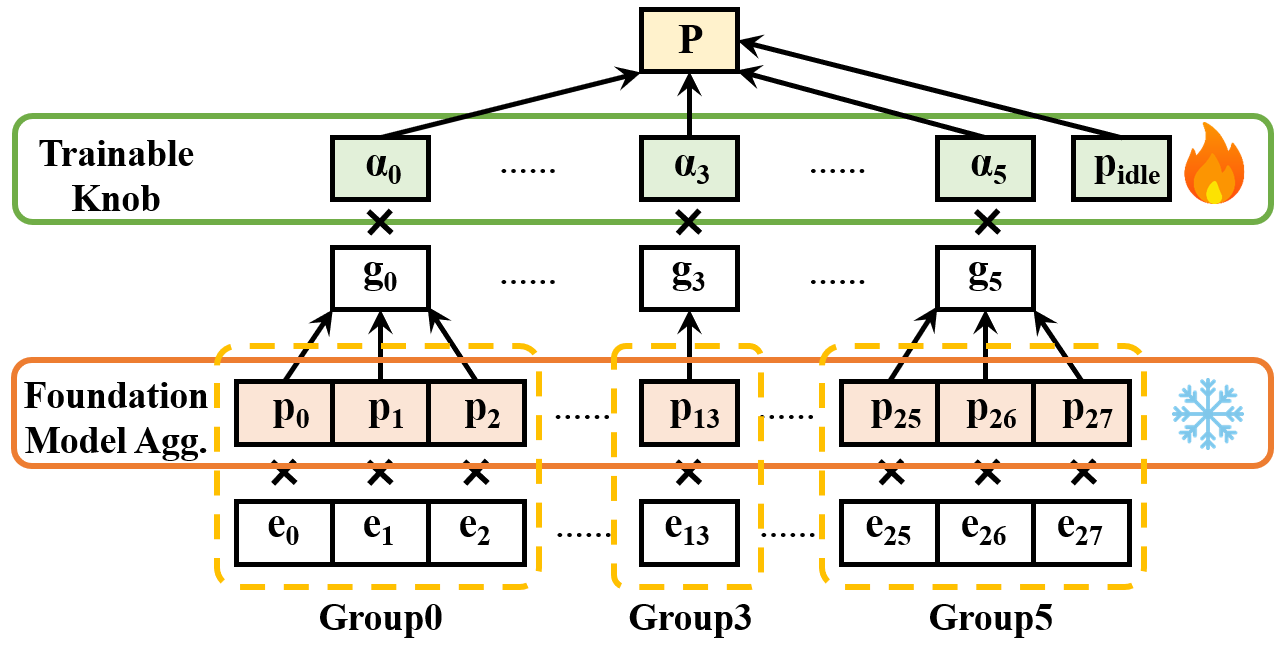}
\vspace{-.2in}
\caption{Illustration of fine-tuning on our target GPU. It introduces a trainable knob for each feature group. When fine-tuning, the given foundation model is frozen, and only the knobs are trained. After fine-tuning, knobs are multiplied by the foundation model to get the final GPU power model.}
\vspace{-.23in}
\label{finetune}
\end{figure}

\subsection{Fine-Tuning on Our Target GPU}
\label{sec:ft}

After the aggregated foundation model is generated, this foundation model is fine-tuned based on the dataset of our target GPU to build the final GPU power model for our target GPU design. The fine-tuning is based on the feature-grouping, where architectural events within the same group are tightly coupled from both the microarchitectural aspect and the data distribution aspect.

\subsubsection{Foundation Fine-Tuning}

We fine-tune the aggregated foundation model on our target GPU based on the feature grouping, which will be introduced in Section~\ref{sec:grouping}. The parameters within the same group should be updated in a coordinated approach because these are tightly coupled. Therefore, we freeze the ratio among parameters within each group during the foundation fine-tuning.

Figure~\ref{finetune} illustrates the details of the foundation fine-tuning. Compared with the original linear model that multiplies $e_i$ and $p_i$ and calculates the sum with an intercept, foundation fine-tuning introduces a trainable knob for each feature group, denoted as $\alpha_j$ in Figure~\ref{finetune}. To be specific, $e_i*p_i$ within each feature group is summed up as $g_j$, and the $g_j$ is then multiplied by $\alpha_j$ and finally summed up with the intercept $p_{\text{idle}}$ to get the final power. During fine-tuning, the foundation model is frozen and only the trainable knob can be updated. In our implementation, we first calculate $g_j$ of each feature group and form a vector of $g_j$. Then, this new vector is adopted as the feature of the linear regression, where the trained coefficients are $\alpha_j$ and the intercept is $p_{\text{idle}}$. After the fine-tuning, these trainable knobs are multiplied by the original foundation model to get the final GPU power model.

\subsubsection{Two-Level Feature Grouping}
\label{sec:grouping}

The two-level feature grouping is performed to partition features into multiple groups that assist the parameter freezing in design fine-tuning. We propose a two-level feature grouping with both microarchitecture-aware grouping and data-driven grouping. 

The microarchitecture-aware grouping partitions features at a coarse granularity, considering the microarchitecture coupling among features. We partition these features, which are architectural events, into three groups based on the domain. As listed in Table~\ref{tbl:event}, the 28 architectural events belong to three domains, including SM array, NoC \& L2Cache, and global memory. The insight is that the microarchitecture within a domain is tightly coupled, so the architectural events are also coupled and then considered to have similar variation trends across GPUs. 

The data-driven grouping further partitions 20 features within SM array domain into a fine granularity. We regard a feature in all samples as the vector of this feature and normalize this vector to a unit vector. Then, the hierarchical clustering, which is feasible for high-dimensional data with few samples, is performed to partition features into multiple groups. The insight is that data-driven grouping identifies features with a similar variation trend across programs and GPUs, regardless of absolute value of each event. 

%% file: _txt/4_result.tex
\section{Experimental Results}

\subsection{Experiment Setup}

We adopt four modern NVIDIA GPUs with Ampere and Ada Lovelace architecture: RTX A6000~\cite{rtxa6000}, RTX 3090~\cite{rtx3090}, RTX 4090~\cite{rtx4090}, and RTX 5880 Ada~\cite{rtx5880}, as listed in Table~\ref{tbl:gpuconfig}. The 65 microbenchmarks are from AccelWattch~\cite{kandiah2021accelwattch}. The real workloads include CUDA Samples 11.0~\cite{cudasample}, Rodinia 3.1~\cite{che2009rodinia}, CUTLASS 1.3~\cite{cutlass}, and Parboil~\cite{stratton2012parboil}. We perform kernel-level evaluation, with multi-kernel benchmarks broken into isolated kernels (Table~\ref{tbl:workloadconfig}).

To collect architectural events, we utilize the NVIDIA Nsight Compute~\cite{ncu}. It measures the real execution statistics when running kernels on the real GPU chips by using the hardware counters. Such an accurate architectural event measurement eliminates the interference of the inaccuracy of the architectural performance simulator, ensuring the evaluation focuses on the power aspect. The architectural events adopted in our experiment are listed in Table~\ref{tbl:event}, including 28 events covering different components. 

To collect power labels, we measure the power from the real GPU chips. In our measurement, we utilize NVIDIA Management Library (NVML)~\cite{nvml} and NVIDIA System Management Interface (nvidia-smi)~\cite{nvidiasmi}. To be specific, we use nvidia-smi to lock the frequency of the SM and memory before power measurement, and then execute an NVML-based monitor program to sample the power consumption during the kernel execution. Similar to the AccelWattch, we heat the GPU to 65$^\circ$C with power-hungry programs before real power measurement. We perform 5 times for each kernel and use the average value as the final power label in our evaluation.

\begin{table}[!t]
      \centering
      \renewcommand{\arraystretch}{1.1}
      \resizebox{0.46\textwidth}{!}{
        \begin{tabular}{ |c|c|c|c| } 
        \hline
        GPU & Tech Node & SM Frequency & Memory Frequency \\
        \hline
        \hline
        NVIDIA RTX A6000 & 8nm & 1410MHz & 2000MHz \\
        \hline
        NVIDIA RTX 3090 & 8nm & 1395MHz & 1219MHz \\
        \hline
        NVIDIA RTX 4090 & 5nm & 2235MHz & 1313MHz \\
        \hline
        NVIDIA RTX 5880 Ada & 5nm & 975MHz & 2250MHz \\
        \hline
        \end{tabular}
        }
        \caption{GPU designs adopted in our evaluation. }
        \vspace{-.25in}
        \label{tbl:gpuconfig}
\end{table}

\begin{table}[!t]
      \centering
      \renewcommand{\arraystretch}{1.1}
      \resizebox{0.46\textwidth}{!}{
        \begin{tabular}{ |c|c||c|c| } 
        \hline
        Kernel Name & Source Benchmark & Kernel Name & Source Benchmark \\
        \hline
        \hline
        \multicolumn{4}{|c|}{CUDA Samples 11.0} \\
        \hline
        cTensor\_K1 & cudaTensorCoreGemm & dct\_K1 & dct8x8 \\
        \hline
        binOpt\_K1 & binomialOptions & dct\_K2 & dct8x8 \\
        \hline
        walsh\_K1 & fastWalshTransform & histo\_K1 & histogram \\
        \hline
        walsh\_K2 & fastWalshTransform & msort\_K1 & mergesort \\
        \hline
        qrng\_K1 & quasirandomGenerator & msort\_K2 & mergesort \\
        \hline
        qrng\_K2 & quasirandomGenerator & sobol\_K1 & SobolQRNG \\
        \hline
        \hline
        
        \multicolumn{4}{|c|}{Rodinia 3.1} \\
        \hline
        bprop\_K1 & backprop & hsopt\_K1 & hotspot \\
        \hline
        bprop\_K2 & backprop & sradv1\_K1 & sradv1 \\
        \hline
        btree\_K1 & b+tree & kmeans\_K1 & kmeans \\
        \hline
        btree\_K2 & b+tree &  &  \\
        \hline
        \hline
        
        \multicolumn{2}{|c||}{CUTLASS 1.3 (cutlass-wmma)} & \multicolumn{2}{c|}{Parboil} \\
        \hline
        cutlass\_K1 & input: 2560x16x2560 & sgemm\_K1 & sgemm \\
        \hline
        cutlass\_K2 & input: 4096x128x4096 & mriq\_K1 & mri-q \\
        \hline
        cutlass\_K3 & input: 2560x512x2560 & sad\_K1 & sad \\
        \hline
        \end{tabular}
        }
        \caption{Real GPU workloads adopted in our evaluation.}
        \vspace{-.25in}
        \label{tbl:workloadconfig}
\end{table}

\subsection{Summary of Baseline Methods}

We adopt two representative existing works as our baselines for comparison. 1) AccelWattch~\cite{kandiah2021accelwattch} is the state-of-the-art architecture-level GPU power model, which is the updated version of GPUWattch \cite{leng2013gpuwattch}. This is built as an analytical power model with power calculation for each component, similar to the McPAT~\cite{li2009mcpat}. A calibration process that adopts a linear function is required before the prediction, training the model on the GPU microbenchmarks of the new target GPU. 2) McPAT-Calib~\cite{zhai2022mcpat} is a representative architecture-level CPU power model that we generalized to the GPU. It builds a machine learning model for power modeling. In our evaluation, the McPAT-Calib is trained on the GPU microbenchmarks of the new target GPU. The machine learning model adopted is XGBoost~\cite{chen2016xgboost}, the best model claimed in McPAT-Calib.

Other recent architecture-level CPU power models, such as PANDA~\cite{zhang2023panda}, FirePower~\cite{zhang2025firepower}, and AutoPower~\cite{zhang2025autopower}, are not included in our evaluation because these works are customized to CPU and can not be generalized to GPU power modeling.

\subsection{Training and Testing Data Setup}

To reflect the real scenario, we adopt one GPU as our target and the remaining three as additional known chips, yielding four testing scenarios (A6000, 3090, 4090, and 5880). The three additional GPUs provide pre-training data with both microbenchmarks and real workloads. The target GPU uses microbenchmarks for fine-tuning and real workloads for testing. Baseline adopts the selected GPU using GPU microbenchmarks for training and real GPU workloads for testing for fair comparison.

\begin{figure}[!t]
\centering
\includegraphics[width=0.47\textwidth]{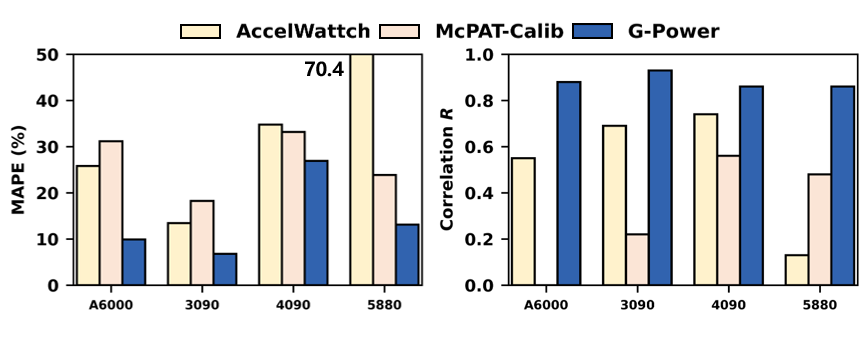}
\vspace{-.25in}
\caption{Accuracy comparison between G-Power and two baselines for four testing scenarios.}
\vspace{-.2in}
\label{expmaper}
\end{figure}

\begin{figure*}[!t]
\centering
\includegraphics[width=0.96\textwidth]{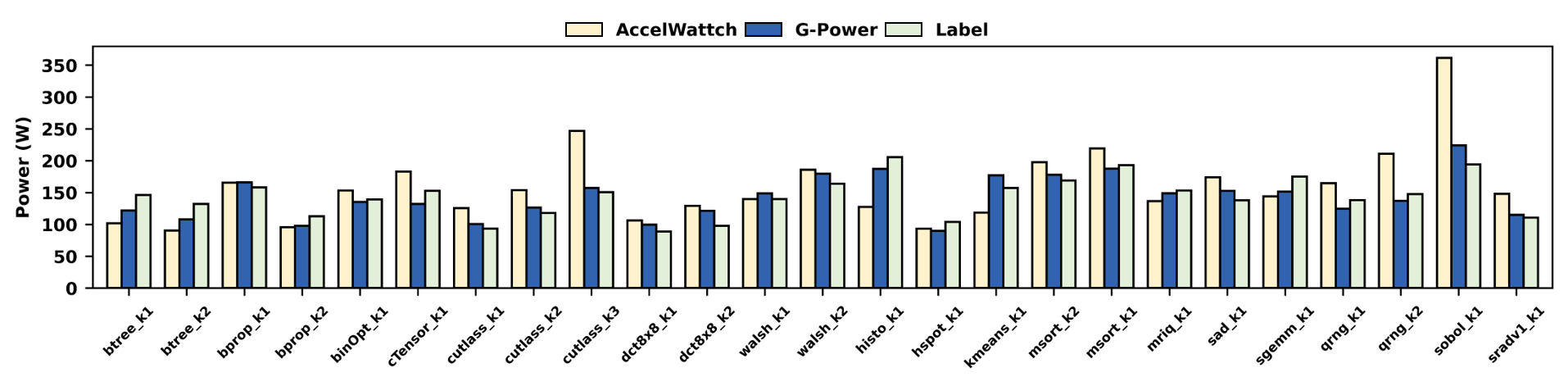}
\vspace{-.25in}
\caption{Detailed RTX A6000 power prediction comparison of AccelWattch and G-Power against power labels.}
\vspace{-.2in}
\label{expworkload}
\end{figure*}

\begin{figure}[!t]
\centering
\hspace{-5mm}
\subfigure[A6000-AccelWattch]{
    \centering
    \includegraphics[height=0.18\textwidth]{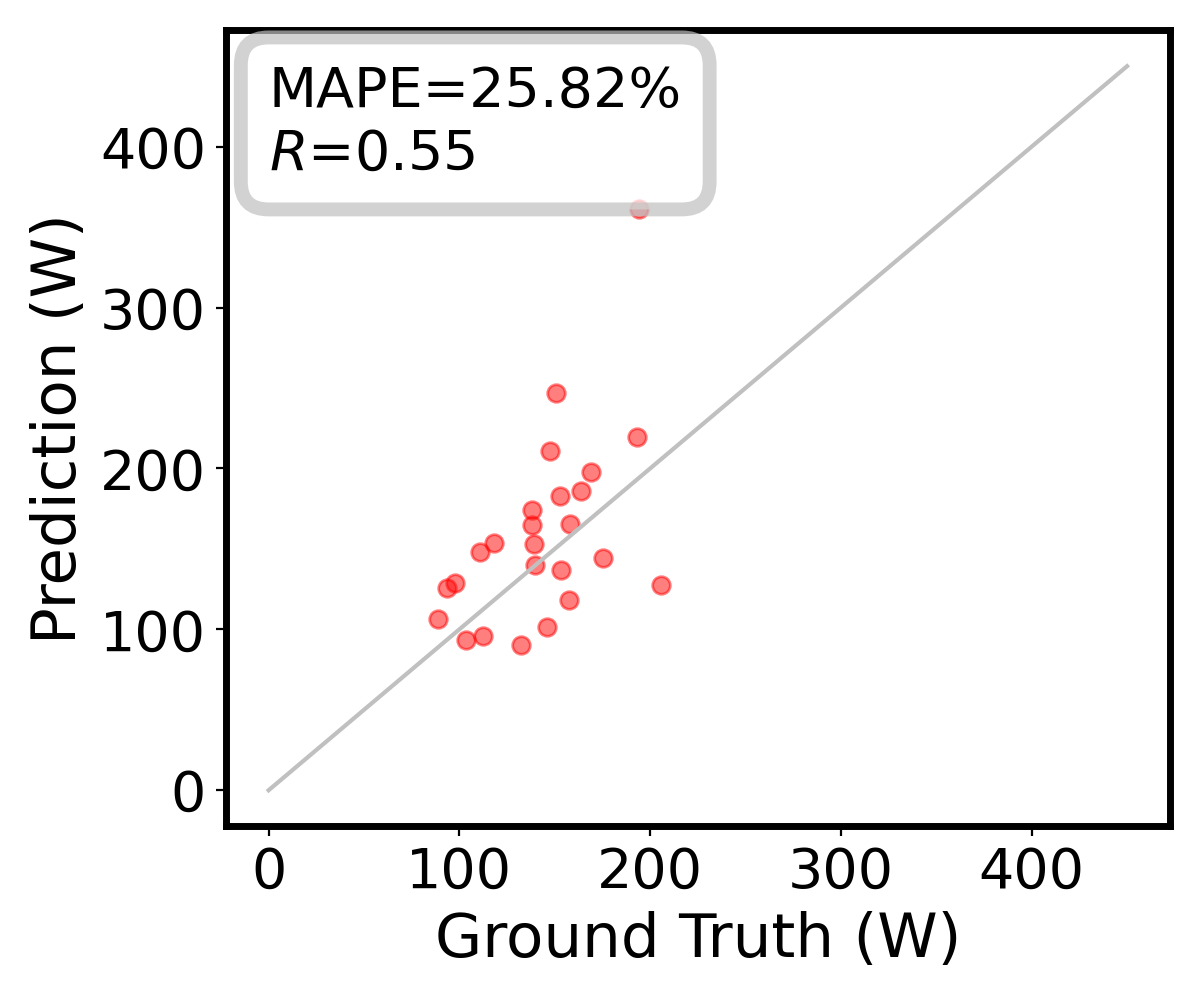}
}
\hspace{-3mm}
\subfigure[A6000-G-Power]{
    \centering
    \includegraphics[height=0.18\textwidth]{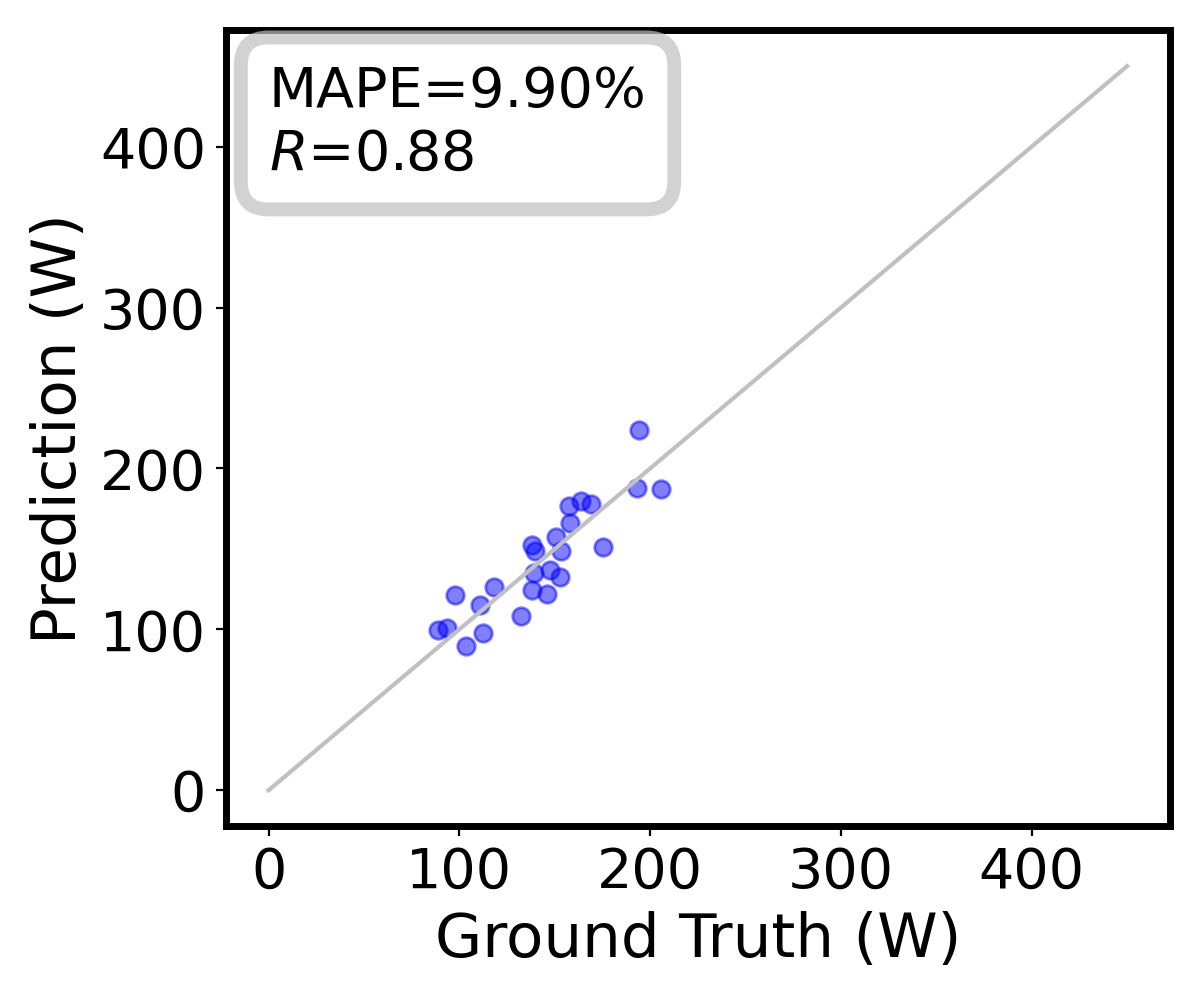}
}

\hspace{-5mm}
\subfigure[3090-AccelWattch]{
    \centering
    \includegraphics[height=0.18\textwidth]{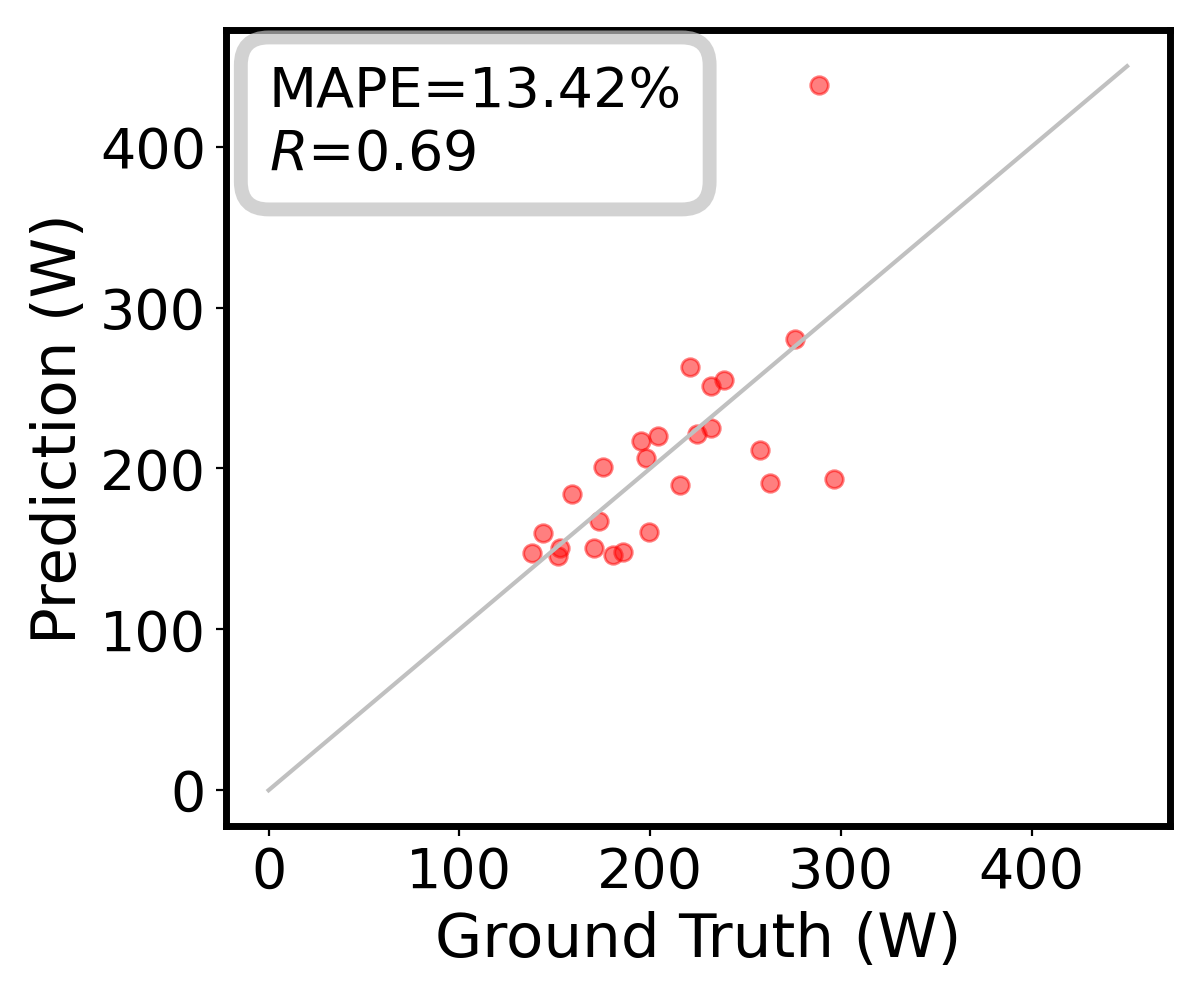}
}
\hspace{-3mm}
\subfigure[3090-G-Power]{
    \centering
    \includegraphics[height=0.18\textwidth]{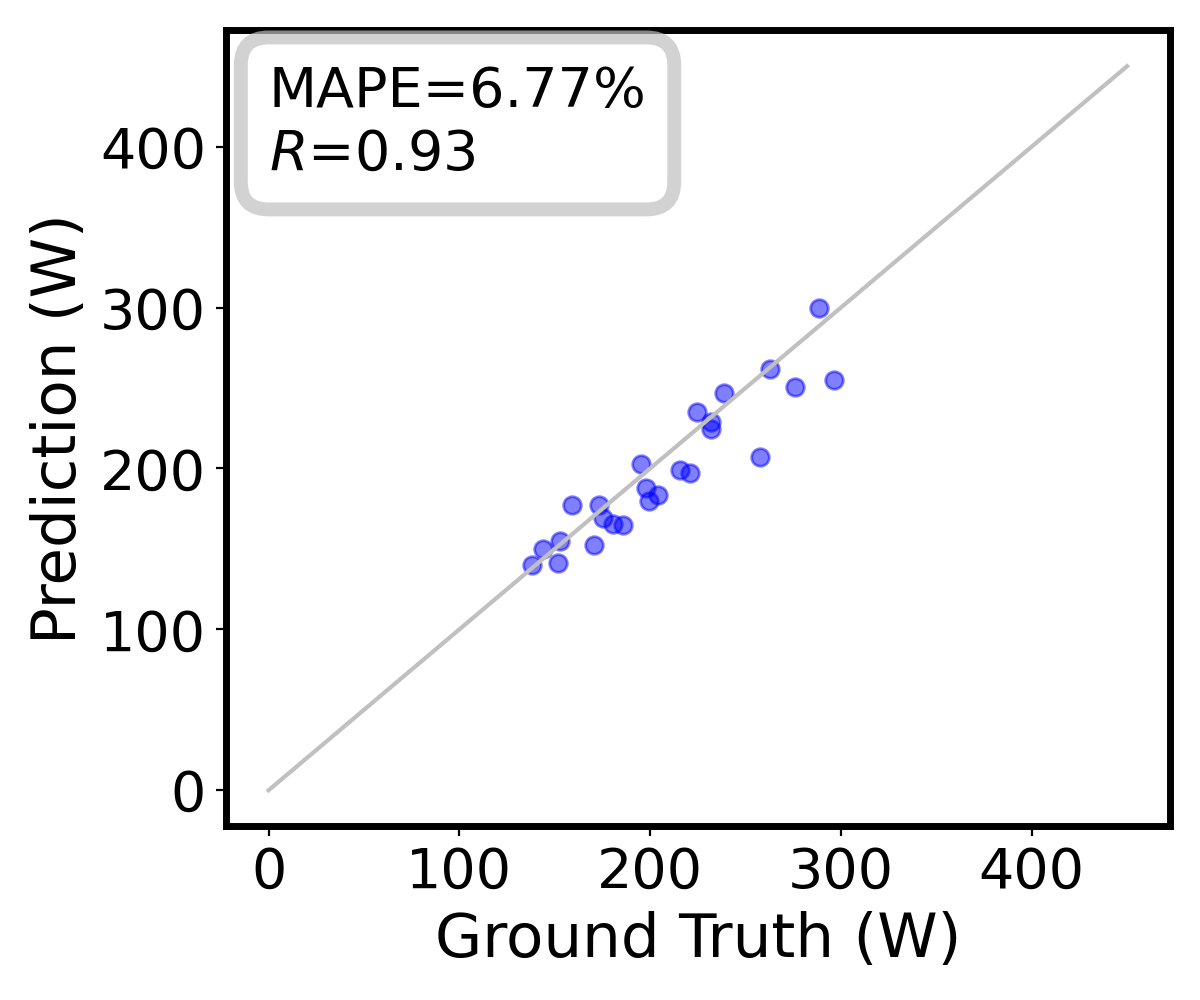}
}

\vspace{-.15in}
\caption{The power prediction visualization for A6000 and 3090 with the AccelWattch and G-Power. }
\vspace{-.2in}
\label{expvisualization}
\end{figure}

\begin{figure}[!t]
\centering
\includegraphics[width=0.46\textwidth]{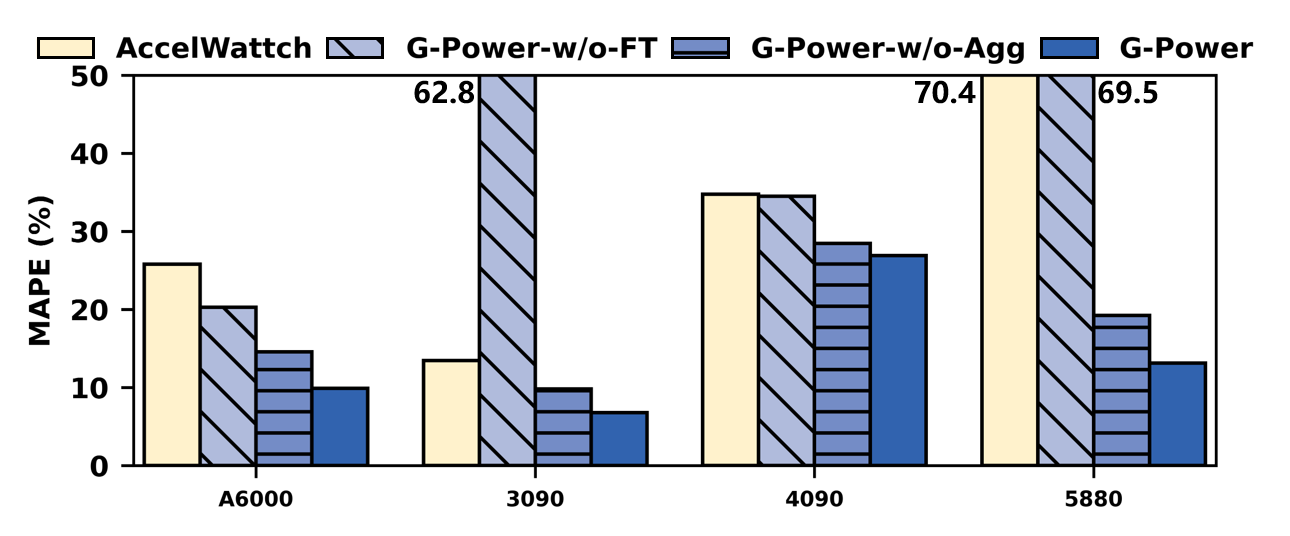}
\vspace{-.18in}
\caption{Ablation studies of G-Power.}
\vspace{-.2in}
\label{expablation}
\end{figure}

\subsection{Power Modeling Accuracy}

Figure~\ref{expmaper} demonstrates the accuracy comparison between our proposed G-Power and our two baselines in four testing scenarios. The accuracy is measured in MAPE and correlation $R$. It shows that our proposed G-Power can significantly outperform all baselines in all testing scenarios. Quantitatively, G-Power can achieve a low MAPE of 14\% and a high correlation $R$ of 0.88 on average. This MAPE is 22\% and 13\% lower than AccelWattch and McPAT-Calib, respectively. The correlation $R$ is 0.36 and 0.57 higher than our two baselines. The superior performance of G-Power over all our baselines demonstrates the effectiveness of utilizing knowledge from additional known GPU chips. 

For AccelWattch, the state-of-the-art architecture-level GPU power model, the prediction on our evaluated four modern GPU designs is inaccurate. The inaccuracy of AccelWattch is due to the limited training data and the data distribution difference between GPU microbenchmarks for training and real GPU workloads for testing. With the foundation that integrates knowledge from multiple additional known GPU chips with comprehensive programs measured, our proposed G-Power can address this data limitation, significantly improving the prediction accuracy. 

Figure~\ref{expworkload} shows the detailed power prediction comparison of AccelWattch and G-Power against the power labels for RTX A6000. It demonstrates that G-Power can achieve consistently higher accuracy over AccelWattch for all our evaluated real GPU workloads. Figure~\ref{expvisualization} visualizes the power prediction for RTX A6000 and 3090 with both the AccelWattch and G-Power. Each point in the figure represents the prediction for a real GPU workload, and the grey line indicates the accurate prediction. This visualization shows that the prediction of G-Power is close to the grey line, while the AccelWattch prediction is far from this line, indicating an advantage of our proposed G-Power. It also demonstrates that the correlation of AccelWattch prediction is dramatically low, which reflects that the power characteristics of different workloads can not be accurately differentiated. In comparison, G-Power achieves a high correlation, correctly identifying power behaviors of different workloads.

For McPAT-Calib, the ML-based power model, its prediction is also inaccurate, especially for the correlation $R$. We observe that the algorithm fits well on the training dataset with GPU microbenchmarks. This indicates that McPAT-Calib has an overfitting problem. The overfitting makes McPAT-Calib can make predictions within a reasonable range similar to the training data, but the relative value is significantly inaccurate, reflected with low correlation $R$.

\subsection{Ablation Studies}

Besides the two baselines discussed above, we also include two methods derived from G-Power for ablation studies: 1) G-Power-w/o-FT. The fine-tuning is not performed, directly adopting the aggregated foundation model as the final power model. 2) G-Power-w/o-Agg. Attention-inspired aggregation is not performed, directly selecting one foundation model for fine-tuning. 

Figure~\ref{expablation} shows the results of our ablation studies, demonstrating the criticality of both fine-tuning and attention-inspired aggregation. The comparison between G-Power-w/o-FT and G-Power indicates that the fine-tuning is critical for G-Power. It is because although different GPU designs have similarities in the power behaviors, the resource amount is different, and there are also differences in microarchitecture design. The comparison between G-Power-w/o-Agg and G-Power shows that directly selecting one foundation model can only achieve sub-optimal accuracy. It is because some GPUs are divergent from our target GPUs, and attention-inspired aggregation can capture the similarity to utilize foundations wisely. 

%% file: _txt/7_conclusion.tex
\section{Conclusion}

We propose G-Power, an architecture-level GPU power modeling framework that utilizes knowledge from additional known GPU chips. 
G-Power adopts a three-phase algorithm to wisely provide foundations with additional known GPU chips and capture the similarity to utilize these foundations for fine-tuning. This foundation-based framework boosts the GPU power modeling accuracy, which is a compelling addition to the GPU architects’ toolbox.

\section*{Acknowledgement}
\vspace{-.02in}
This work is supported by National Natural Science Foundation of China (NSFC) 62304192, Hong Kong Research Grants Council (RGC) CRF-YCRG C6003-24Y. It was partially conducted by ACCESS – AI Chip Center for Emerging Smart Systems, supported by the InnoHK initiative of the Innovation and Technology Commission of the Hong Kong Special Administrative Region Government.   
\vspace{-.05in}